\documentclass[pra,10pt,twocolumn,groupedaddress,floatfix,showpacs]{revtex4-2}
\usepackage[utf8]{inputenc}
\usepackage[T1]{fontenc}     
\usepackage[british]{babel}  
\usepackage[sc,osf]{mathpazo}
\usepackage[scaled=0.86]{berasans}  
\usepackage[colorlinks=true, allcolors=blue, urlcolor=blue]{hyperref}  
\usepackage{graphicx} 
\usepackage[babel]{microtype}  
\usepackage{amsmath,amssymb,amsthm,bm,amsfonts,mathrsfs,bbm} 

\usepackage{xspace}  
\usepackage{pgfplots}
\usepackage{xcolor,colortbl}
\usepackage{array}
\usepackage{bigstrut}

\newcounter{subeq}
\renewcommand{\thesubeq}{\theequation\alph{subeq}}
\newcommand{\newsubeqblock}{\setcounter{subeq}{0}\refstepcounter{equation}}
\newcommand{\mysubeq}{\refstepcounter{subeq}\tag{\thesubeq}}

\newcommand{\be}{\begin{equation}}
\newcommand{\ee}{\end{equation}}
\newcommand{\bea}{\begin{eqnarray}}
\newcommand{\eea}{\end{eqnarray}}
\newcommand{\bes}{\begin{equation*}}
\newcommand{\ees}{\end{equation*}}
\newcommand{\beas}{\begin{eqnarray*}}
	\newcommand{\eeas}{\end{eqnarray*}}

\newtheorem{thm}{Theorem}
\newtheorem*{thm*}{Theorem}

\newtheorem*{lem*}{Lemma}

\newtheorem*{lipschitzLem*}{Lemma \ref{lipschitz}}
\newtheorem*{lipschitzCubeLem*}{Lemma \ref{lipschitzCube}}
\newtheorem*{pgmNearlyOptimalThm*}{Theorem \ref{pgmNearlyOptimal}}

\begin{document}

\title{Device-independent bounds from Cabello's nonlocality argument}


\author{Ashutosh Rai$^{1,~2}$}
\email{ashutosh.rai@kaist.ac.kr}

\author{Matej Pivoluska $^{ 1,~3,~ 4}$}
\email{pivoluskamatej@gmail.com}

\author{Martin Plesch $^{1,~3}$}
\email{martin.plesch@savba.sk}
\affiliation{$^{1}$Institute of Physics, Slovak Academy of Sciences, 845 11 Bratislava, Slovakia,}
\affiliation{$^{2}$School of Electrical Engineering, Korea Advanced Institute of Science and Technology (KAIST), 291 Daehak-ro, Yuseong-gu, Daejeon 34141, Republic of Korea,}
\affiliation{$^{3}$Institute of Computer Science, Masaryk University, 602 00 Brno, Czech Republic,}
\affiliation{$^4$ Institute for Quantum Optics and Quantum Information - IQOQI Vienna, Austrian Academy of Sciences, Boltzmanngasse 3, 1090 Vienna, Austria}

\author{Souradeep Sasmal}
\email{souradeep.007@gmail.com}
\affiliation{Light and Matter Physics, Raman Research Institute, Bengaluru 560080, India}

\author{Manik Banik}
\email{manik.banik@iisertvm.ac.in}
\affiliation{School of Physics, IISER Thiruvananthapuram, Vithura, Kerala 695551, India}

\author{Sibasish Ghosh}
\email{sibasish@imsc.res.in}
\affiliation{Optics \& Quantum Information Group, The Institute of Mathematical Sciences, HBNI, C. I. T. Campus, Taramani, Chennai 600113, India}


\begin{abstract}

Hardy-type arguments manifest Bell nonlocality in one of the simplest possible ways. Except for demonstrating nonclassical signature of entangled states in question, they can also serve for device-independent self-testing of states, as shown, e.g., in \href{https://journals.aps.org/prl/abstract/10.1103/PhysRevLett.109.180401}{Phys. Rev. Lett. 109, 180401 (2012)}. Here we develop and broaden these results to an extended version of Hardy's argument, often referred to as Cabello's nonlocality argument. We show that, as in the simpler case of Hardy's nonlocality argument, the maximum quantum value for Cabello's nonlocality is achieved by a pure two-qubit state and projective measurements that are unique up to local isometries. We also examine the properties of a more realistic case when small errors in the ideal constraints are accepted within the probabilities obtained and prove that also in this case the two-qubit state and measurements are sufficient for obtaining the maximum quantum violation of the classical bound.

\end{abstract}


\maketitle


\section{Introduction}
Over the years Bell inequalities have evolved from study in the foundations of quantum mechanics~\cite{Bell 1964, Brunner et al 2014, Scarani 2019} to invaluable resource for many applications in quantum information science~\cite{Barrett 2005, Pironio et al 2016, Acin and Navascues 2017, Pivoluska and Plesch 2015, Kar and Banik 2016}. The key finding from violations of Bell inequalities is that, on sharing entangled quantum states, it is possible to generate nonlocal correlations between parties that are spacelike separated.

A fundamental aspect is to learn about the maximum possible violation of Bell inequalities in quantum mechanics; these are referred to as device-independent quantum bounds~\cite{Pironio et al 2016}. One such bound was first derived by Cirel'son~\cite{Cirel'son 1980} for the paradigmatic Bell-CHSH inequality~\cite{CHSH}. The quantum bounds on violation of Bell inequalities are termed as device independent because on considering state and measurements as black box~\cite{Acin and Navascues 2017}, solely from the collected experimental statistics of a corresponding Bell-type experiment, one can extract very useful information about the actual quantum state and measurements. Moreover, sometimes from such experimental statistics alone, one can learn (up to isometry) even the exact quantum state and (or) measurements inside the black box; this is a topic of research known as self-tests of quantum state and measurements~\cite{Supic and Bowles 2020, Coladangelo et al}. More broadly, deriving device independent bounds for Bell inequalities leads to knowing more about the geometry of the set of quantum correlation~\cite{NPA, Pitowski 2008, Goh et al 2018, Duarte et al 2018, Ishizaka 2018, Rai et al 2018, Escola et al 2020} as these bounds are achieved on the boundary of the quantum set~\cite{NPA, Goh et al 2018}.

 For certifying nonlocal correlations, one can also provide nonlocality arguments of the type introduced by Hardy \cite{Hardy 1993}. The Hardy-type demonstrations of nonlocality are simple (often referred to as the simplest demonstration of Bell nonlocality \cite{Mermin 1995}) and yet they can reveal rich structures in the quantum set of correlations \cite{Seshadreesan and Ghosh 2011, Rabelo et al 2012, Ahanj et al 2010, Gazi et al 2010, Das et al 2013a, Das et al 2013b, Mancinska et al 2014}. Recently, such results derived for Hardy and Hardy-type correlations have been shown to be useful for witnessing postquantum correlations \cite{Das et al 2013b}, for constructing a device-independent dimension witness \cite{Mukherjee et al 2015}, and for devising quantum random number generators \cite{Li et al 2015, Ramnathan et al, Sasmal et al}. 
 
 One of the characteristic features of Hardy-type nonlocality arguments is that, unlike standard Bell inequalities, they follow from certain prior constraints placed on some of the outcome probabilities. Geometrically it means that, while studying Hardy or Hardy-type correlations, we study the quantum set (and its boundary) in the cross sections defined by the underlying constraints. Then a natural question to ask is what are the device-independent bounds on nonlocality demonstrated by Hardy or Hardy like arguments? Along this direction Rabelo, Zhi and Scarani \cite{Rabelo et al 2012} have derived the device-independent bound for Hardy's nonlocality argument, as well as some more interesting results about self-testing of the maximal Hardy state, and an experimentally implementable nonideal version of Hardy's tests. In this work, we ask and try to answer such questions for a Hardy-type nonlocality argument often referred to as Cabello's nonlocality argument \cite{Cabello 2002, Liang and Li 2005, Kunkri et al 2006}, which is a generalization of Hardy's argument with fewer constraints on outcome probabilities. We derive the optimal degree of success (device-independent bound) for Cabello's test of nonlocality. Then we prove that the state(s) leading to the device-independent bound are self-testable up to the application of local isometries. Finally, we analyze an experimentally feasible nonideal version of Cabello's test and show that, similar to the ideal Cabello's test of nonlocality, pure qubit state and projective measurements are sufficient for a nonideal test. 
 
 The remaining sections of the paper are organized along these lines. In the Sec.~\ref{CNA} we first introduce Cabello's nonlocality argument. Then, in Sec.~\ref{qubit-CNA} we characterize all pure two-qubit states that can pass Cabello's test of nonlocality. In Sec.~\ref{DIB-CNA} we derive the device-independent bound on the degree of success for Cabello's nonlocality argument and in section~\ref{ST-CNA} we show that quantum state(s) giving the device-independent bound can be self-tested. Next, in Sec.~\ref{NI-CNA} we modify the ideal Cabello's test to an experimentally realizable nonideal version and derive the results and its implementation. Finally, we summarize our work in the concluding Sec.~\ref{Conclusions}.

\section{Cabello's nonlocality argument} \label{CNA}

Cabello's nonlocality argument \cite{Liang and Li 2005, Kunkri et al 2006} can be summarized as follows. Suppose two parties, Alice and Bob, share parts of a physical system. On her part, Alice can choose to perform one or the other measurement $x\in \{A_0,A_1\}$, whereas Bob on his part has a similar choice of measuring $y\in\{B_0,B_1\}$. Let any choice of measurement by Alice and Bob have binary outcomes, say $a\in\{\pm1\}$ for Alice and $b\in\{\pm1\}$ for Bob. Then a Cabello's nonlocality argument constitutes four joint probabilities with two of them constrained to take the value zero as follows:
\begin{align}
&~P(+,~+~\vert~ A_0,~B_0)\equiv q, \label{eq1} \\
\newsubeqblock
\mysubeq          &~P(+,~-~\vert~ A_1,~B_0)=0, \label{eq2a} \\
\mysubeq          &~P(-,~+~\vert~ A_0,~B_1)=0,\label{eq2b} \\
         &~P(+,~+~\vert~ A_1,~B_1)\equiv p.\label{eq3}
\end{align}
In the above, note that there are only two equality constraints, those given by Eq.~(\ref{eq2a}) and Eq.~(\ref{eq2b}), whereas $p$ and $q$, appearing respectively in Eq.~(\ref{eq1}) and Eq.~(\ref{eq3}), are only given names to the joint probabilities and they can take arbitrary possible values. Now, the statement of Cabello's nonlocality argument is that if, for some outcome probability distribution along with satisfying the two equality constraints, the condition $p>q$ is also true, then it cannot be described by any local realistic theory. A proof of the statement is following. Suppose there is some set of deterministic (realistic) local hidden variables $\Lambda$ such that one can write $$P(a,b\vert x,y)=\sum_{\lambda\in\Lambda}~\mbox{pr}(\lambda)~x(\lambda)~y(\lambda),$$ 
i.e., for any hidden variable $\lambda \in \Lambda$ the outcome $x(\lambda)~[y(\lambda)]$ of local observable $x$~[$y$] takes a definite value from $\{\pm1\}$ [here $\mbox{pr}(\lambda)$ is the probability distribution of hidden variables $\lambda$]. From condition $p>q$ we have $p>0$, and then Eq.~(\ref{eq3}) implies that there is a nonempty subset $S_{\Lambda}\subset \Lambda$ such that, for all $\lambda \in S_{\Lambda}$, ~$A_1(\lambda)=+1$ and $B_1(\lambda)=+1$. Further, for all $\lambda \in S_{\Lambda}$, from Eq.~(\ref{eq2b}) we get $A_0(\lambda)=+1$, and  similarly Eq.~(\ref{eq2a}) gives $B_0(\lambda)=+1$. Therefore, $P(+1,+1\vert A_0, B_0, S_{\Lambda})=P(+1,+1\vert A_1,B_1)=p$. However, $S_{\Lambda}\subset \Lambda$ implies $P(+1,+1\vert A_0,B_0,S_{\Lambda})\leq P(+1,+1\vert A_0,B_0)$, i.e. $p\leq q$; this contradicts the assumption $p>q$. Thus one can conclude that any correlation satisfying condition $p>q$, along with the two constraint equations, must be nonlocal. The degree of success for Cabello's nonlocality argument can then be defined as
\begin{equation}
\mathcal{S}=p-q>0. \label{eq4}
\end{equation}
To sum up, when constraint Eqs.~(\ref{eq2a}) and (\ref{eq2b}) are satisfied: on observing some value $\mathcal{S}\leq 0$ the resulting correlation is local; on the other hand, we start to witness nonlocal correlations when we find that $\mathcal{S}> 0$. Here we note that Cabello's nonlocality argument is a generalization of the nonlocality argument by Hardy which constitutes three constraints \cite{Hardy 1993, Rabelo et al 2012}, i.e., in a corresponding Hardy's test, along with constraint Eqs.~(\ref{eq2a}) and (\ref{eq2b}), additionally the probability in Eq.~(\ref{eq1}) is also constrained as $P(+1,+1\vert A_0,B_0)\equiv q=0$, and nonlocality is certified when $p>0$.

\section{Two qubit states showing Cabello's nonlocality} \label{qubit-CNA}

In order to provide examples and to characterize the two-qubit state and measurements which can lead to Cabello's nonlocality, let us consider the following general pure two-qubit state shared between two parties Alice and Bob:
\begin{eqnarray*}
&~&\vert\Psi\rangle_{AB}=c_{00}~\vert 00\rangle+c_{01}~\vert 01\rangle+c_{10}~\vert 10\rangle+c_{11}~\vert 11\rangle, \\
&~&\mbox{such~that}~~\sum_{i,j\in \{0,1\}}\vert c_{ij}\vert^2=1, \nonumber
\end{eqnarray*}
and let the (projective) measurement of Alice be $x=|u^+_{x}\rangle \langle u^+_{x}\vert-|u^-_x\rangle \langle u^-_x\vert$, and that of Bob be $y=|v^+_y\rangle \langle v^+_y\vert-|v^-_y\rangle \langle v^-_y\vert$. One can choose the basis for the measurements $x \in \{A_0,A_1\}$ and $y\in \{B_0,B_1\}$ as follows
\begin{align}
\newsubeqblock
\mysubeq A_0\equiv\label{eq5a}
\left\{\!\begin{aligned}|u^+_{A_0}\rangle &=|0\rangle,\\\nonumber
|u^-_{A_0}\rangle&=|1\rangle,\end{aligned}\right\},  \\[4pt]
\mysubeq A_1\equiv\label{eq5b}
\left\{\!\begin{aligned}|u^+_{A_1}\rangle &=\cos\left(\tfrac{\alpha}{2}\right)\vert 0\rangle+e^{i \phi}\sin\left(\tfrac{\alpha}{2}\right)\vert 1\rangle,\\\nonumber
|u^-_{A_1}\rangle&=-\sin\left(\tfrac{\alpha}{2}\right)\vert 0\rangle+e^{i \phi}\cos\left(\tfrac{\alpha}{2}\right)\vert 1\rangle,\end{aligned}\right\}, \\[4pt]
\mysubeq B_0\equiv\label{eq5c}
\left\{\!\begin{aligned}|v^+_{B_0}\rangle&=|0\rangle,\\\nonumber
|v^-_{B_0}\rangle&=|1\rangle,\end{aligned}\right\},  \\[4pt]
\mysubeq B_1\equiv\label{eq5d}
\left\{\!\begin{aligned}|v^+_{B_1}\rangle&=\cos \left(\tfrac{\beta}{2}\right)\vert 0\rangle + e^{i \xi}\sin \left(\tfrac{\beta}{2}\right)\vert 1\rangle,\\\nonumber
|v^-_{B_1}\rangle&=-\sin \left(\tfrac{\beta}{2}\right)\vert 0\rangle + e^{i \xi}\cos \left(\tfrac{\beta}{2}\right)\vert 1\rangle,\end{aligned}\right\},
\end{align}
where $0<\alpha,\beta<\pi$ and $0\leq \phi,\xi<2\pi$. We remark that there is no loss of generality by fixing observable $A_0=B_0=\sigma_z\equiv|0\rangle \langle 0\vert-|1\rangle \langle 1\vert$ provided we keep the other two observables $A_1$ and $B_1$, and the state $\vert\Psi\rangle_{AB}$, in the most general form. This is due to the following invariance property for measurement statistics
\begin{align*}
P(a,b\vert x,y)&=\big\vert \langle\tilde{\Psi}\vert \tilde{u}^a_x\otimes \tilde{v}^b_y\rangle\big\vert^2\\
&=\big\vert \langle\tilde{\Psi}\vert U^{\dagger}U\vert \tilde{u}^a_x\otimes \tilde{v}^b_y\rangle\big\vert^2\\
&=\big\vert \langle\Psi\vert u^a_x\otimes v^b_y\rangle\big\vert^2,
\end{align*}
where $U=U_A\otimes U_B$ is some (product) unitary transformation such that $U_A(U_B)$ act on the local state space of Alice (Bob) and vectors $\vert \tilde{u}^a_x\rangle$ ($\vert\tilde{v}^b_y\rangle$) define some arbitrary projective measurement by Alice (Bob). In particular, on applying unitary defined by a map $U\vert \tilde{u}^a_{A_0}\otimes \tilde{v}^b_{B_0}\rangle\mapsto \vert \frac{1-a}{2}\rangle \otimes \vert \frac{1-b}{2}\rangle$ where $a,b\in\{\pm1\}$, we get a general pure state $\vert \Psi\rangle=U\vert\tilde{\Psi}\rangle$ and observables of the form given by Eqs.~(\ref{eq5a})-(\ref{eq5d}).

From the state and measurements in the considered canonical form, all the two-qubit pure states respecting the constraints Eqs.~(\ref{eq2a}) and (\ref{eq2b}) in Cabello's test must satisfy the following two orthogonality conditions:
\begin{align}
\newsubeqblock
\mysubeq \vert\Psi\rangle_{AB} &\perp |u^+_{A_1}\rangle \otimes |1\rangle, \label{eq6a}\\
\mysubeq \vert\Psi\rangle_{AB} &\perp |1\rangle \otimes |v^+_{B_1}\rangle. \label{eq6b}
\end{align}
On imposing the above two conditions, it turns out that the class of all possible pure two qubit states (up to multiplication by some global phase) is of the form:

\begin{align}
\vert\Psi^{\mathcal{C}}\rangle_{AB}
&\!=e^{i \delta } \sqrt{1\!-\!c^2 \left\{\!1\!+\!\tan ^2\!\left(\tfrac{\text{$\alpha $}}{2}\right)\!+\!\tan
	^2\!\left(\tfrac{\text{$\beta $}}{2}\right)\!\right\}}\vert 00\rangle \nonumber \\
&\!-c\left\{e^{-i \text{$\phi $}} \tan\!
\left(\tfrac{\text{$\alpha $}}{2}\right)\vert 01\rangle + e^{-i \text{$\xi $}} \tan\!\left(\tfrac{\text{$\beta
		$}}{2}\right)\vert 10\rangle \right\}\nonumber \\
	&\!+c\vert 11\rangle, \label{eq7}
\end{align}
where parameters $c$ and $\delta$ satisfy $0\leq c\leq1$ and $0\leq \delta < 2\pi$. Then, computing the degree of success for Cabello's nonlocality argument yields
\begin{align}\nonumber
\mathcal{S}&^{qubits} =\frac{1}{4} \Big[\cos (\text{$\alpha $})\!\cos (\text{$\beta $})+\cos (\text{$\alpha $})+\cos (\text{$\beta $})-3 \Big. \nonumber\\
&\Big.\!\!-\!2c \sin(\text{$\alpha $})\!\sin(\text{$\beta $})\!\cos(\delta\!
+\!\text{$\xi $}\!+\!\text{$\phi $})\!\sqrt{\!{\scriptstyle c^2\!\left(\!-\!\tan^2\!\left(\tfrac{\!\text{$\alpha $}}{2}\!\right)\!\right)}{\scriptstyle-}\tfrac{2 c^2}{\cos (\text{$\beta $})+1}{\scriptstyle +1}}\!\Big. \nonumber\\\nonumber
&\Big.\!\!+\!2c^2\!\left(\!\cos(\text{$\alpha $})\!
(\cos(\text{$\beta $})\!-\!1)\!+\!2\tan^2\!\left(\!\tfrac{\text{$\alpha $}}{2}\!\right)\!\right.\\
&\quad\quad\quad\quad\quad\quad\quad\quad\!-\!\Big.\left.\cos(\text{$\beta$})\!+\!2\tan^2\!\left(\!\tfrac{\text{$\beta $}}{2}\!\right)\!\!+\!1\!\right)\!\quad\Big]. \label{eq8}
\end{align}

There are many values of state and measurement parameters for which $\mathcal{S}^{qubits}>0$ and thus certify quantum nonlocality through Cabello's nonlocality argument; for instance, if state parameters take the value $\delta=0,~ c=\sqrt{9/15}$ and the measurement parameters take the values $\alpha=\beta=\pi/3,~\phi=\xi=\pi/2$, then $\mathcal{S}^{qubits}= 3/80>0$.

\subsection*{Maximum degree of success for two qubit states}
Now we like to maximize the degree of success for Cabello's nonlocality argument over all two qubit states and two outcome positive-operator-valued-measure (POVM) measurements. First we note that since any mixed state can be written as a convex mixture of pure states, it is sufficient to perform optimization over set of all pure two qubit state. Next we further note that any two outcome POVM measurement on a qubit can be always implemented as a classical mixture of two outcome projective measurements \cite{D'Ariano et al}. One can see this as follows, a two outcome POVM measurement, say $M\equiv\{E, \mathbb{I}-E\}$, on a qubit can be represented as a set of two positive semi-definite operators acting on a Hilbert space $\mathbb{C}^2$ such that sum of the two operators is $\mathbb{I}$. In order to satisfy the positive semi-definiteness conditions, operator $E$ takes the following form:
\begin{eqnarray}
&E&=a_0 \mathbb{I} + \eta ~\hat{a}\cdot \sigma,~~\mbox{such~that}\nonumber \\
&0&\leq a_0\leq 1~~\mbox{and}~~0\leq \eta \leq \min~\{a_0,~1-a_0\}, \nonumber
\end{eqnarray}
where $\hat{a}$ is a unit vector in $\mathbb{R}^3$ and $\sigma=(\sigma_x,\sigma_y,\sigma_z)$ is the vector of three Pauli matrices. Now on considering the three projective measurements, $M_1\equiv\{\mathbb{I}, 0\}$, $M_2\equiv\{0, \mathbb{I}\}$, and $M_3\equiv\{\frac{1}{2}(\mathbb{I}+\hat{a}\cdot \sigma),~ \frac{1}{2}(\mathbb{I}-\hat{a}\cdot \sigma)\}$, one can easily verify that $$M=(a_0-\eta)~M_1+(1-a_0-\eta)~M_2 + 2\eta~M_3.$$
Thus, POVM measurement $M$ can be implemented as a classical mixture of three projective measurements $M_1$, $M_2$, and $M_3$. Therefore, we can now say that it is sufficient to perform the optimization over all projective measurements and pure two qubit states, i.e., an optimization of the expression derived in Eq.~(\ref{eq8}) over all the parameters $\alpha,\beta,\phi~\mbox{and}~\xi$. On performing the optimization we find that the maximum possible value of the Cabello's nonlocality argument over qubit states is as follows
\begin{align}
\mathcal{S}_{max}^{qubits}&= \frac{1}{3} \left(\tfrac{\sqrt[3]{\left(307+39 \sqrt{78}\right)^{2}}-29}
{\sqrt[3]{307+39 \sqrt{78}}}-5\right)
\approxeq 0.1078, \label{eq9}\\
\mbox{when}~ &\mbox{parameters take the following values} \nonumber \\
\newsubeqblock
\mysubeq \alpha  \label{eq10a}
&=2 \mbox{tan}^{-1}\sqrt{\tfrac{\left(\sqrt[3]{359-12 \sqrt{78}}+\sqrt[3]{359+12 \sqrt{78}}\right)-1}{12}},\\
\mysubeq\beta&=\alpha \approxeq 1.6136,\label{eq10b}\\
\mysubeq \delta&=\pi-\phi-\xi, \label{eq10c}\\[2 pt]
 \mysubeq c&= \frac{1}{6} \left(\tfrac{\sqrt[3]{\left(307+39 \sqrt{78}\right)^{2}}-29}
{\sqrt[3]{307+39 \sqrt{78}}}-2\right)
\approxeq 0.5539.\label{eq10d}
\end{align}
The pure two qubit state which gives the maximum value then, on substituting Eqs.~(\ref{eq10a}-\ref{eq10d}), is of the form
\begin{align}
\newsubeqblock\nonumber
\vert\Psi^{\mathcal{C}}_{max}\rangle\!_{AB}\!&=\kappa_{00}~e^{-i (\text{$\xi $}+\text{$\phi $})}\vert 00\rangle\\
\mysubeq&~~~+\!\kappa_{01}\!\left\{\!e^{-i \text{$\phi $}}\vert 01\rangle\!+\!e^{-i \text{$\xi $}}\vert 10\rangle\!\right\}\!+\!\kappa_{11}\vert 11\rangle, \label{eq11a}\\[10pt]
~\mbox{where}, \nonumber \\
\mysubeq \kappa_{00}&=~\frac{1}{6} \left({\scriptstyle 4-}\tfrac{\sqrt[3]{(53-6 \sqrt{78})^2}+1}{\sqrt[3]{53-6 \sqrt{78}}}\right)\approxeq -0.1573, \label{eq11b}\\
 \kappa_{01}&= \!\!\left(\frac{12-\frac{31\times 6^{2/3}}{\sqrt[3]{67 \sqrt{78}-414}}+\sqrt[3]{6 \left(67 \sqrt{78}-414\right)}}{12}\right)^{\frac{1}{2}} \nonumber \\
\mysubeq&\approxeq\!-0.5781, \label{eq11c}\\
\mysubeq \kappa_{11}&=\!\!\frac{1}{6} \left(\tfrac{\sqrt[3]{\left(307+39 \sqrt{78}\right)^2}-29} {\sqrt[3]{307+39 \sqrt{78}}}-2\right)\approxeq 0.5539. \label{eq11d}
\end{align}
For achieving the maximum value, the two parameters $\phi$ and $\xi$ can be assigned any real values, however, each of the two parameter should take same value in the quantum state Eq.~(\ref{eq11a}) and measurements Eqs.~(\ref{eq5a}-\ref{eq5d}); in addition, the value of parameters $\alpha$ and $\beta$ appearing in the measurements is given by Eq.~(\ref{eq10a}-\ref{eq10b}). Thus, we note here that for any arbitrarily fixed value of parameters $\phi$ and $\xi$, the two qubit state which gives the maximum success is unique. Moreover, all the two qubit states giving the optimal value form an equivalence class under local unitary operations.

\section{Device independent bound on Cabello's nonlocality} \label{DIB-CNA}
In this section, we will show that the maximum success of Cabello's test over all two qubit states is in fact also the optimal value that can be achieved with bipartite quantum states of any (finite) dimension.

\begin{thm}
The maximum possible value for the degree of success in Cabello's test of nonlocality over all bipartite quantum states of finite dimensions can be achieved just with projective measurements on a two qubit pure state, i.e., $S^{Q}_{max}=S_{max}^{qubits}$.
\end{thm}

{\bf Proof:} The proof follows by applying similar analysis as in Refs.~\cite{Rabelo et al 2012,Banik et al 2013}. Say a general bipartite state $\rho$ is shared between Alice and Bob. Let $\Pi_{a\vert x}$ be the measurement operator associated with outcome $a$ when Alice measures observable $x$. Similarly, $\Pi_{b\vert y}$ denotes the measurement operator associated with outcome $b$ when Bob measures observable $y$. Then the joint probability of getting outcomes $(a,b)$ for measurements $(x,y)$ is
\begin{equation*}
P(a,b\vert x,y)=\mbox{Tr}(\rho~ \Pi_{a\vert x} \otimes \Pi_{b\vert y}).
\end{equation*}
Since there is no restriction on dimension, by applying the Neumark's dilation theorem, we consider only projective measurements. The observable of Alice and Bob are then hermitian operators, with eigenvalues $\pm1$, which can be expressed as,
\begin{eqnarray*}
x&=& (+1)~\Pi_{+\vert x} ~+~(-1)~\Pi_{-\vert x}~~~~\mbox{where}~x\in\{A_0,A_1\}\\
y&=& (+1)~\Pi_{+\vert y} ~+~(-1)~\Pi_{-\vert y}~~~~\mbox{where}~y\in\{B_0,B_1\}
\end{eqnarray*}
Note that Alice's observable constitutes of two pairs of projection operators $\{\Pi_{+\vert A_0},~\Pi_{-\vert A_0}\}$ and $\{\Pi_{+\vert A_1},~\Pi_{-\vert A_1}\}$, which satisfy $\Pi_{+\vert A_0} +\Pi_{-\vert A_0}=I$ and $\Pi_{+\vert A_1} +\Pi_{-\vert A_1}=I$, and similarly for Bob. So we can use a lemma proved in Ref.~\cite{Masanes 2006}. The lemma essentially states that for four projection operators, lets say~$\Pi_{+\vert \mathcal{O}_0}$, ~$\Pi_{-\vert \mathcal{O}_0}$, ~$\Pi_{+\vert \mathcal{O}_1}$, and ~$\Pi_{-\vert \mathcal{O}_1}$, which act on a Hilbert space $\mathcal{H}$ and satisfy conditions $\Pi_{+\vert \mathcal{O}_0} +\Pi_{-\vert \mathcal{O}_0}=I$ and $\Pi_{+\vert \mathcal{O}_1} +\Pi_{-\vert \mathcal{O}_1}=I$, there is an orthonormal basis in $\mathcal{H}$ where all the four projectors are simultaneously block diagonal with each block of the size either $2\times 2$ or $1\times 1$. Then such a basis induces a direct sum decomposition of the Hilbert space as $\mathcal{H}=\oplus_s\mathcal{H}^s$, where dimension of each component subspace $\mathcal{H}^s$ is at most two, and all the four projection operators have a decomposition $\Pi_{\pm\vert \mathcal{O}_{0(1)}}=\oplus_s~\Pi_{\pm\vert \mathcal{O}_{0(1)}}^s$ such that $\Pi_{\pm\vert \mathcal{O}_{0(1)}}^s$ acts (as non identity) only on subspace $\mathcal{H}^s$. The projector on subspace $\mathcal{H}^s$ can be written as $\Pi^s=\Pi_{+\vert \mathcal{O}_{0}}^s+\Pi_{-\vert \mathcal{O}_{0}}^s=\Pi_{+\vert \mathcal{O}_{1}}^s+\Pi_{-\vert \mathcal{O}_{1}}^s$.

The stated lemma when applied to observable $\{A_0,A_1\}$ of Alice which induce a decomposition $\mathcal{H}_A=\oplus_i\mathcal{H}_A^i$, and observable $\{B_0,B_1\}$ of Bob which induce a decomposition $\mathcal{H}_B=\oplus_j\mathcal{H}_B^j$, gives
\begin{align}
\newsubeqblock
\mysubeq P(a,b\vert x,y)&=\sum_{i,j} \mu_{ij}~\mbox{Tr}(\rho_{ij}~ \Pi^i_{a\vert x} \otimes \Pi^j_{b\vert y}),\label{eq12a}\\
\mysubeq &\equiv \sum_{i,j} \mu_{ij}~P_{ij}(a,b\vert x,y), \label{eq12b}
\end{align}
 where $\mu_{ij}=\mbox{Tr}(\rho~\Pi^i\otimes\Pi^j)$ and $\rho_{ij}=\frac{(\Pi^i\otimes\Pi^j~\rho~\Pi^i\otimes\Pi^j)}{\mu_{ij}}$. Notice that $\mu_{ij}\geq 0$ for all $i,j$ and $\sum_{i,j}\mu_{ij}=1$, and $\rho_{ij}$ is a trace one positive operator acting on subspace of types, $\mathbb{C}\otimes \mathbb{C},~\mathbb{C}\otimes \mathbb{C}^2,~\mathbb{C}^2\otimes \mathbb{C},~\mathbb{C}^2\otimes \mathbb{C}^2$.

Now the concluding argument of the proof is as follows. If the joint probability $P(a,b\vert x,y)$ satisfy the two constraints Eq.~(\ref{eq2a}) and Eq.~(\ref{eq2b}) in the Cabello's test, then it follows from Eq.~(\ref{eq12b}) that the joint probability $P_{ij}(a,b\vert x,y)$ generated from the subspace $\mathcal{H}_A^i\otimes\mathcal{H}_B^j$ will also satisfy the constraint equations. Moreover, the degree of success for Cabello's nonlocality argument can be expressed as
\begin{eqnarray}
\mathcal{S}=p-q&=&\sum_{i,j}\mu_{ij}~(p_{ij}-q_{ij})\nonumber \\
&\leq& \underset{i,j}{max}~(p_{ij}-q_{ij})= \mathcal{S}_{max}^{qubits}.\label{eq13}
\end{eqnarray}
$\mbox{Here}~p_{ij}=P_{ij}(+,+\vert A_1,B_1)~\mbox{and}~q_{ij}=P_{ij}(+,+\vert A_0,B_0)$. Thus, from Eq.~(\ref{eq13}) it follows that the maximum value of $\mathcal{S}$ over any finite dimensions should be upper bounded by the maximum value achieved with two qubit states and projective measurements, i.e., $\mathcal{S}^{qudits}_{max}\leq \mathcal{S}_{max}^{qubits}$. However, we also have the opposite inequality  $\mathcal{S}^{qudits}_{max}\geq \mathcal{S}_{max}^{qubits}$, which leads us to conclude that $\mathcal{S}^{qudits}_{max}= \mathcal{S}_{max}^{qubits}$.

\section{Self-test of state leading to maximal Cabello's nonlocality} \label{ST-CNA}
From the result derived in the previous section, we can now prove a self-test result for the pure two qubit state $\vert\Psi^{\mathcal{C}}_{max}\rangle_{AB}$ which gives the maximum degree of success for the Cabello's nonlocality argument. In light of the proof for device independent bound that is given for Cabello's test, it is intuitive that if the value $S^{Q}_{max}=S_{max}^{qubits}$ is achieved in an experiment with quantum system of unknown (finite) dimension, then state of the system must be direct sum of copies of $\vert\Psi^{\mathcal{C}}_{max}\rangle_{AB}$. In what follows we formally state and prove such a result.

\begin{thm}
If the degree of success $S^{Q}_{max}$ is observed in a Cabello's test of nonlocality on measuring some unknown quantum state $\vert\chi \rangle_{AB}$, then the state of unknown system  is equivalent up to local isometries to  $\vert \sigma\rangle_{AB}\otimes\vert\Psi^{\mathcal{C}}_{max}\rangle_{A'B'}$, where $\vert\Psi^{\mathcal{C}}_{max}\rangle$ is given by Eqs.~(\ref{eq11a}-\ref{eq11d}) and $\vert \sigma\rangle$ is an arbitrary bipartite state.
\end{thm}

{\bf Proof:} The theorem is proved following the same argument as in Ref.~\cite{Rabelo et al 2012}. Observable $A_0$ and $B_0$ can be chosen such that both of them have their eigenstates  $\{\vert0\rangle,\vert1\rangle,\vert2\rangle,...\}$ (i.e., vectors of the computational basis, where the length of basis set is determined from the local dimension of Hilbert space on which the observable act). Note that this does not affect the general analysis, provided we keep the other observable $A_1$, $B_1$ and the state $\vert\chi\rangle$ in most general representation. Then corresponding to observable $A_0$ and $B_0$, the decomposed projectors on $\mathcal{H}_A^i$ and $\mathcal{H}_B^j $ subspace can be respectively expressed as
\begin{eqnarray}
\Pi^i_{+\vert A_0}&=&\vert2i\rangle\langle2i\vert, ~~~~\Pi^i_{-\vert A_0}=\vert2i+1\rangle\langle2i+1\vert,\nonumber \\
\Pi^j_{+\vert B_0}&=&\vert2j\rangle\langle2j\vert, ~~~~\Pi^j_{-\vert B_0}=\vert2j+1\rangle\langle2j+1\vert \nonumber,
\end{eqnarray}
where $i,j\in\{0,1,2,...\}$. Now, the degree of success of Cabello's test $p_{ij}-q_{ij}$ in $\mathcal{H}_A^i\otimes \mathcal{H}_B^j $ subspace can take the value $\mathcal{S}^{Q}_{max}$ if and only if $\rho_{ij}=\vert\Psi^{\mathcal{C}}_{max}\rangle_{ij}\langle\Psi^{\mathcal{C}}_{max}\vert$, where $\vert\Psi^{\mathcal{C}}_{max}\rangle_{ij}$ is the two qubit state given by Eqs.~(\ref{eq11a}-\ref{eq11d}). Therefore, the unknown state $\vert\chi \rangle$ can give the maximal value of, $p-q=\sum_{i,j} \mu_{ij}~(p_{ij}-q_{ij})$, the success for Cabello's test, if and only if
\begin{equation}
\vert\chi \rangle=\bigoplus_{i,j}\sqrt{\mu_{ij}}~ \vert\Psi^{\mathcal{C}}_{max}\rangle_{ij}.
\end{equation}
We note that here the parameters $\phi$ and $\xi$ for state $\vert\Psi^{\mathcal{C}}_{max}\rangle_{ij}$ are independent of the indices $i,j$, since they are determined uniquely by the measurement operators which on achieving the maximal value takes the same form in all subspace $\mathcal{H}_A^i$ and $\mathcal{H}_B^j $. Finally, we provide a local isometries $\Phi_A$ and $\Phi_B$ such that
\begin{equation}
(\Phi_A\otimes\Phi_B)\vert~\chi\rangle_{AB}\vert 00\rangle_{A'B'}=\vert \sigma\rangle_{AB}\otimes\vert\Psi^{\mathcal{C}}_{max}\rangle_{A'B'},
\end{equation}
where components of the $\vert 00\rangle_{A'B'}$ are local ancilla qubits appended to the unknown state $\vert~\chi\rangle_{AB}$, and after application of the local isometry $\Phi_A\otimes\Phi_B$ we like to get the target state $\vert\Psi^{\mathcal{C}}_{max}\rangle_{A'B'}$ along with some bipartite junk state $\vert \sigma\rangle_{AB}$. Such an isometry map is as follows
 \begin{align}
 &~\Phi_A=\Phi_B=\Phi, \nonumber \\[5pt]
 \newsubeqblock
\mysubeq &~\Phi~\vert 2k,0\rangle_{XX'} \mapsto \vert 2k,0\rangle_{XX'},\\
\mysubeq &~\Phi~\vert 2k+1,0\rangle_{XX'} \mapsto \vert 2k,1\rangle_{XX'},\\[2pt]
 &~\mbox{where}~XX'\in\{AA',BB'\}.\nonumber
 \end{align}
 This concludes our proof.

 \section{Cabello's test for non-ideal constraints} \label{NI-CNA}
 The constraints appearing in an ideal Cabello's test demands that two of the joint probabilities in the test should be zero. This may be very difficult to insure in any real experiment. In experiments, a more realistic constraint can be of the form
 \begin{align}
 \newsubeqblock
 \mysubeq          &~P(+,~-~\vert~ A_1,~B_0)\leq \varepsilon, \label{eq17a} \\
 \mysubeq          &~P(-,~+~\vert~ A_0,~B_1)\leq \varepsilon,\label{eq17b}
 \end{align}
 where $\varepsilon \geq 0$ is some small error bound. Note that on choosing different error bounds say $\varepsilon_1$ and $\varepsilon_2$ respectively in Eqs.~(\ref{eq17a}) and (\ref{eq17b}), one can always define a same error bound for both the probabilities as $\varepsilon= \max\{\varepsilon_1,\varepsilon_2\}$. With considered error bound on the constrained probabilities, i.e. from Eqs.~(\ref{eq17a}-\ref{eq17b}), the local bound on degree of success in a non-ideal Cabello's test takes the form

 \begin{equation}
 \mathcal{S}=P(+,+\vert A_1,B_1)-P(+,+\vert A_0,B_0) ~\leq ~2\varepsilon. \label{eq18}
 \end{equation}
 Note that Eq.~(\ref{eq18}) follows on applying the (realistic) constraints to the locality bound given by the CH inequality \cite{CH, Rabelo et al 2012}. In what follows we show that for a sufficiently large values of the error bound $\varepsilon$, the maximum degree of success $\mathcal{S}$ can still be achieved with pure two qubit states and projective measurements on them. Since the analytical technique used to show such a result in ideal test is not applicable to the non-ideal scenario, we will show this numerically. For that, first we derive a quantum upper bound on the maximum possible value of $\mathcal{S}$ (under the given constraints) by applying the well known tool developed by Navascues, Pironio, and Acin (NPA) \cite{NPA} implementable through a MATLAB program NPA Hierarchy \cite{qetlab}. Then, we derive a quantum lower bound, for the same, by maximizing over certain class of pure two qubit states and projective measurements. We find that for sufficiently large values of the error bound $\varepsilon$, to a very high order of accuracy, the obtained lower and upper bounds coincide.

\begin{figure}[h!]
 	 	\begin{center}
 	 		\includegraphics[angle=0, width=0.48\textwidth]{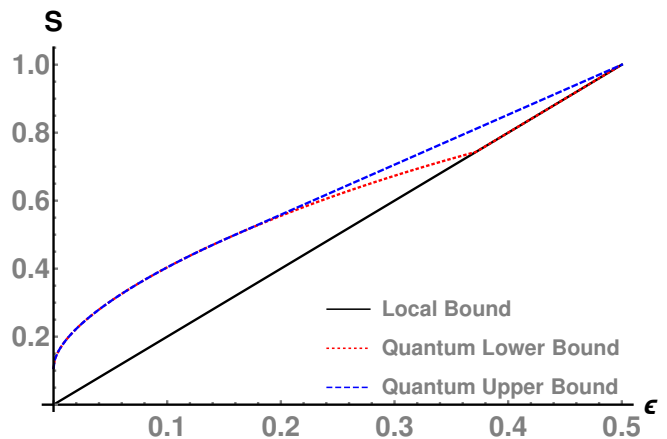}
 	 		\caption{ The figure shows plots for local bound (solid line), a quantum lower bound (dotted line), and quantum upper bound (dashed line) for the maximum degree of success $\mathcal{S}$ of non-ideal Cabello's test of nonlocality, for different values of error bounds $0\leq \varepsilon\leq 1/2$. The range of $\varepsilon$ where quantum lower and upper bound coincide gives the device independent bound for the non-ideal Cabello's test.}  \label{fig}
 	 	\end{center}	
 	 \end{figure}

 First we note that the quantity $\mathcal{S}$ being a difference of two probabilities can not exceed the value $1$, therefore, the local bound $2 \varepsilon$ is trivially respected for the error bound $\varepsilon\geq1/2$. So we need to consider only the error bounds in the range $0\leq \varepsilon < 1/2$. Then, to obtain quantum upper bounds, we maximized $\mathcal{S}$ under the constraints given by Eqs.~(\ref{eq17a}) and (\ref{eq17b}), and the over all the points in $Q^3$ (the third level of the NPA Hierarchy). We choose the NPA level $Q^3$ because, in general, compared to the lower NPA levels $Q^1,Q^{1+ab}, Q^2$ the higher level $Q^3$ should give a tighter upper bound (with an increase in computational cost which is not too high). Note that from the NPA criteria $Q\subseteq Q^3$, therefore we obtain a quantum upper bound. Next, for deriving quantum lower bounds we considered the following class of pure two qubit states,

 	\begin{align}
 \vert\psi\rangle_{AB}&=s_{00}~e^{-i (\text{$\xi $}\!+\!\text{$\phi $})}\vert 00\rangle \nonumber \\
 	&~~~+\!s_{01}\left\{\!e^{-i \text{$\phi $}}\vert 01\rangle\!+\!e^{-i \text{$\xi $}}\vert 10\rangle\!\right\}\!+\!s_{11}\vert 11\rangle, \label{eq19} 
 	\end{align}
 	 and projective measurements defined by Eqs.~(\ref{eq5a}-\ref{eq5d}). We then maximized $\mathcal{S}$ numerically over all the state and measurement parameters under the constraints given by Eqs.~(\ref{eq17a}) and (\ref{eq17b}), and obtain quantum lower bounds for different values of error bounds $0\leq \varepsilon\leq 1/2$. The quantum upper and lower bounds and the locality bounds are plotted in the Fig.(\ref{fig}). We find that the quantum upper and lower bounds coincide with an order of accuracy at least $10^{-7}$ in the error bound range $0\leq \varepsilon \lessapprox 0.158 $. Thus even in a non-ideal Cabello's test, with a sufficient margin for errors, pure qubit state and projective measurements give the optimal quantum value of $\mathcal{S}$.
 	 
 To further investigate the gap between quantum lower and upper bounds in the parameter range $0.158 \lessapprox \varepsilon < 1/2 $, we have also considered the most general qubit states and projective measurements. We have seen that the gap remained the same, showing that the Ansatz chosen in Eq.~(\ref{eq19}) is optimal. Considering that, in the simplest Bell scenario, at the third level of NPA hierarchy we have $Q^3 \approxeq Q$ (to a very high numerical precision) \cite{NPA}, and the fact that any two outcome POVM on a qubit can be simulated from two outcome projective measurements \cite{D'Ariano et al}, the gap between lower and upper bound can potentially be closed only by considering, at least for one party, a higher than two dimensional quantum system.

\section{Conclusion} \label{Conclusions}

To summarize, in this paper, for the simplest Bell scenario, we studied a generalized Hardy-type nonlocality argument known as Cabello's nonlocality argument \cite{Li et al 2015, Kunkri et al 2006}. We derived the device-independent bound for the degree of success of Cabello's test of nonlocality and proved that it can be achieved with a pure two-qubit state and projective measurements. Further, we showed that the two-qubit pure state giving the maximum success can be self-tested in an ideal experiment. Finally, we showed that even in a nonideal Cabello's test, i.e., when constraint probabilities has some nonzero error bound, device-independent bounds for the degree of success are saturated by a class of pure qubit states and projective measurements, over a considerable range for error bounds. Our results are a natural extension of the results derived in \cite{Rabelo et al 2012} for Hardy's test of nonlocality. Along with the foundational relevance of the derived results, the nonideal version of Cabello's test can be implementable in future experiments.

\begin{acknowledgments}
A.R., M. Pivoluska, and M. Plesch acknowledge funding and support from VEGA Project No. 2/0136/19. M. Pivoluska and M. Plesch additionally acknowledge GAMU Project No. MUNI/G/1596/2019. M.B. acknowledges research grant through the INSPIRE-faculty fellowship from the Department of Science and Technology, Government of India.
\end{acknowledgments}

\end{document}